\documentclass[12pt,epsfig]{article} 
\usepackage{epsfig}
\begin{document}

\begin{center}

STUDY OF THE REACTION $np \rightarrow np \pi^+ \pi^-$ AT 
       INTERMEDIATE ENERGIES 
\\
\vspace{0.3cm}

A.P. Jerusalimov, Yu.A.Troyan, A.Yu.Troyan, A.V.Belyaev, E.B.Plekhanov
\\

JINR, Dubna, Moscow region, 141980, Russia 

\end{center}

\vspace{0.3cm}
\begin{abstract}
  The reaction $np \rightarrow np \pi^+ \pi^-$ was studied at the various
momenta of incident neutrons. It was shown that the characteristics of the
reaction at the momenta above 3 GeV/c could be described by the model of reggeized
$\pi$~exchange (OPER). At the momenta below 3 GeV/c, it was necessary to use additionally 
the mechanism of one baryon exchange (OBE).   
\end{abstract}

\vspace{1.0cm}
\section{Introduction: study of inelastic np interactions  
                  at  accelerator facility of  LHEP JINR}

  The data about inelastic np interactions were obtained due to irradiation
of 1m hydrogen bubble chamber (4$\pi$ geometry) by quasimonochromatic neutron 
beam ($\delta P < 2.5\%$) at the following incident momenta:\\ 
$P_0$=1.25, 1.43, 1.73, 2.23, 3.10, 3.83, 4.10 and 5.20 GeV/c

 The unique of fullness and precision data are obtained~\cite{np_inel}. 
It permits to carry out the detailed study of inelastic $np$ interactions in a
wide region of energies. 

\begin{figure}[h]
\includegraphics[width=1.0\textwidth]{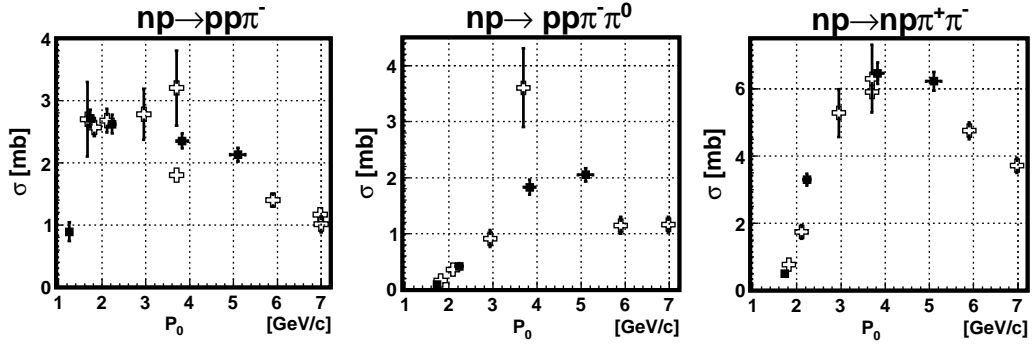}
\caption{Cross-sections of some inelastic np interactions
         (black squares - our data)}
\label{fig:Fig1}
\end{figure}

\section{The reaction $np \rightarrow np \pi^+ \pi^-$ at $P_0$ $>$ 3 GeV/c}

  This reaction is characterized by:\\
\hspace*{1.5cm} - plentiful  production  of  the $\Delta$-resonance (see Fig.2),\\
\begin{figure}[h]
\includegraphics[width=1.05\textwidth]{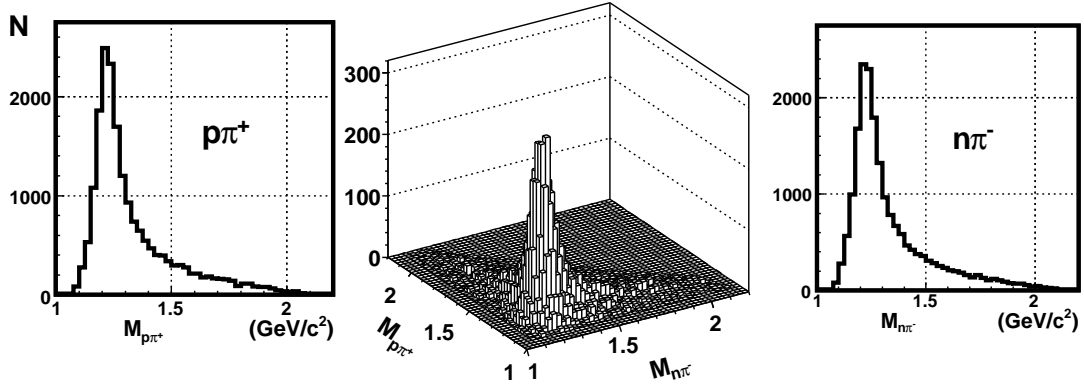}
\caption{The distributions of $M_{p\pi^+}$ and $M_{n\pi^-}$ from the reaction
  $np \rightarrow np \pi^+ \pi^-$ at $P_0$ = 3 GeV/c}
\label{fig:2}
\end{figure}
\hspace*{1.4cm} - large peripherality of the secondary nucleons.

 To study the mechanism of the reaction it was chosen the model of
reggeized $\pi$~exchange (OPER), developed in ITEP~\cite{OPER}.
       
 The advantages of OPER~model are: \\ 
\hspace*{1.5cm} - small number of free parameters (3  in our case),\\
\hspace*{1.5cm} - wide region of the described energies (2$\div$200 GeV),\\
\hspace*{1.5cm} - calculated values are automatically normalized to the reaction
 cross-section.\\

The following main diagrams correspond to the reaction $np \rightarrow np \pi^+ \pi^-$ within the 
framework of  OPER model:
\begin{figure}[h]
\includegraphics[width=1.0\textwidth]{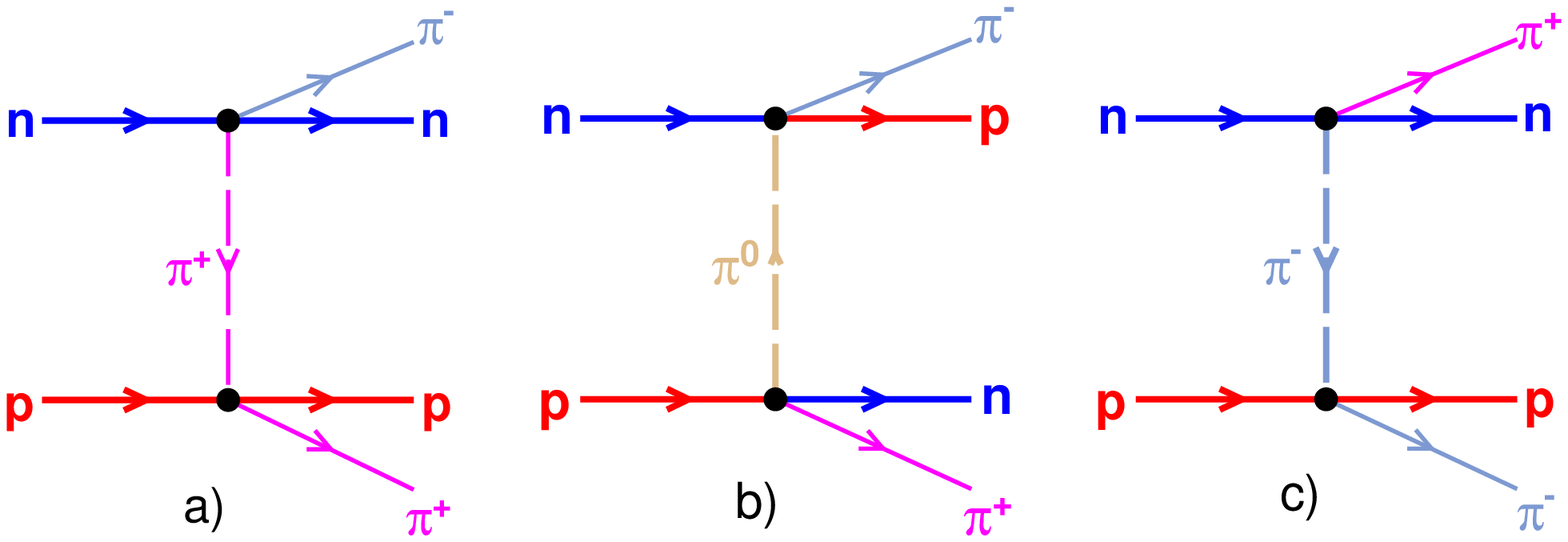}
\caption{OPER diagrams 2$\times$2 for the reaction $np \rightarrow np \pi^+ \pi^-$}
\label{fig:Fig3}
\end{figure}

Matrix element for the diagrams a, b and c from Fig.~\ref{fig:Fig3} is written in the 
following form:
 $$M_1= T_{\pi N \rightarrow \pi N}·F_2·T_{\pi N \rightarrow \pi N} /(t - m_{\pi}^2),$$
 where  $T_{\pi N \rightarrow \pi N}$  - amplitude of elastic $\pi N \rightarrow \pi N$ scattering 
 off mass shell,\\
\hspace*{1.0cm} $F_2$ - form-factor, going away off mass shell of $T_{\pi N \rightarrow \pi N}$ 
amplitudes,\\
\hspace*{1.0cm} $1/(t - m_{\pi}^2)$  - $\pi$-meson propagator.

  The data of elastic $\pi N \rightarrow \pi N$ were taken from PWA~\cite{PWA}. 

  The analysis shows, that interference between diagrams 3a ,3b and 3c is negligible~\cite{INTERF}.

The study has shown that it is not necessary to take into account the contribution of 
the  "hanged" diagrams (Fig.)into the reaction cross-sections at $P_0$ $<$ 10 GeV/:

\begin{figure}[h]
\includegraphics[width=0.80\textwidth]{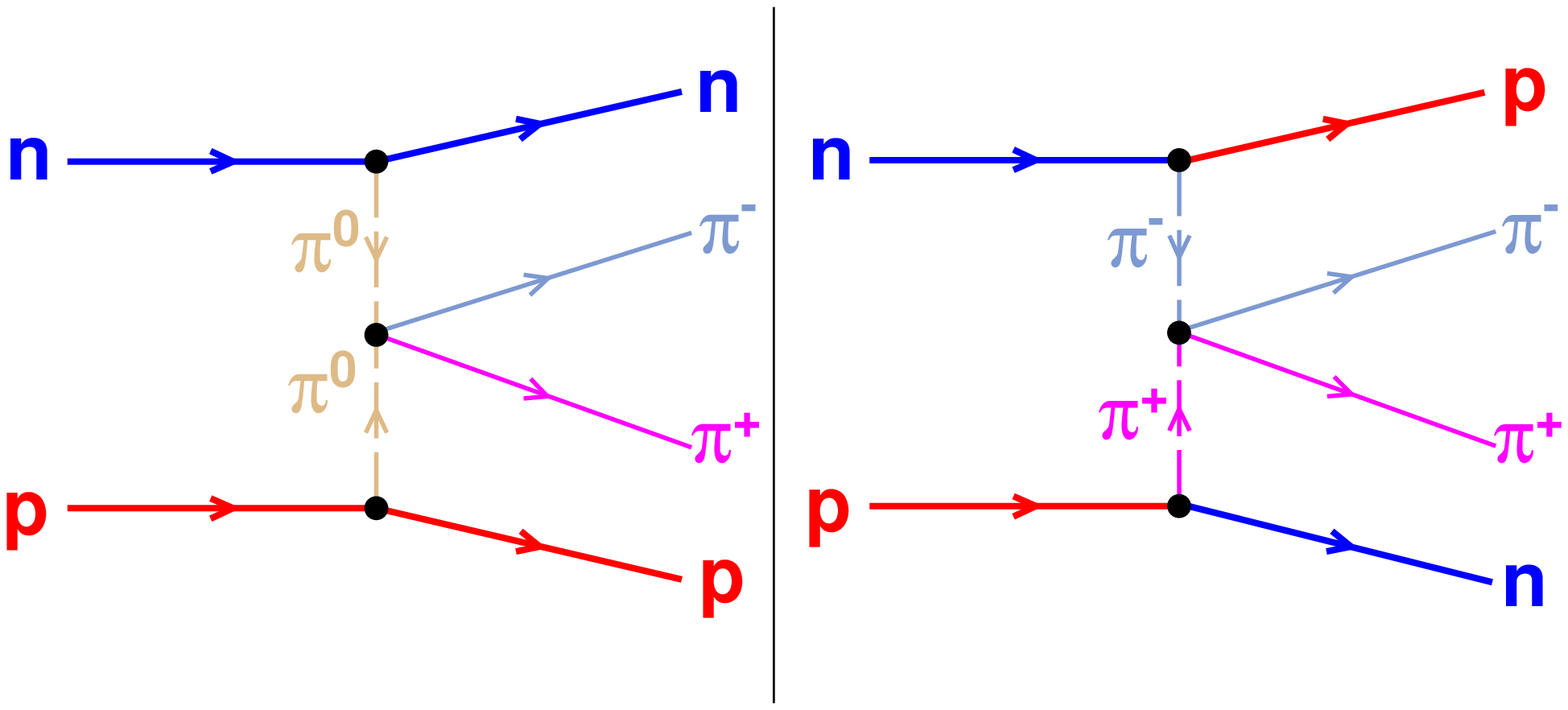}
\caption{"Hanged" OPER diagram for the reaction $np \rightarrow np \pi^+ \pi^-$}
\label{fig:Fig4}
\end{figure}

\vspace{0.5cm}
 It was shown in~\cite{pbarp} that the use of some specific cuts permits to select the kinematic
region of the reaction $np \rightarrow np \pi^+ \pi^-$ in which the contribution of the diagrams
3a, 3b and 3c consists up to 95 $\%$ at $P_0$ $>$ 3 GeV/c.

   Fig.5 shows some distributions for the reaction $np \rightarrow np \pi^+ \pi^-$ for 
this region at $P_0$=5.20 GeV/c (solid curves - results of calculations using OPER~model).
\begin{figure}[h]
\includegraphics[width=0.85\textwidth]{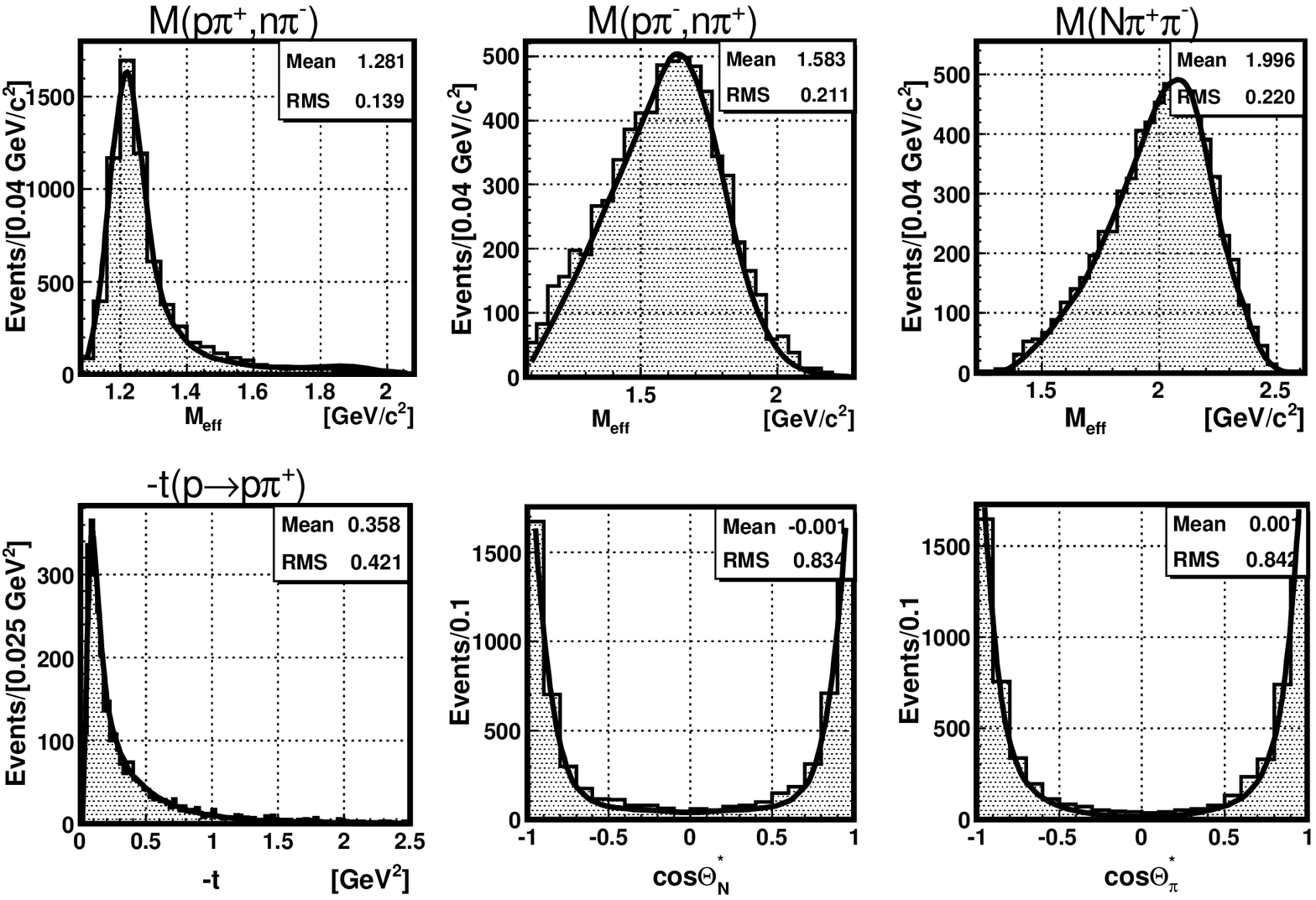}
\caption{Distributions for the reaction $np \rightarrow np \pi^+ \pi^-$ at $P_0$=5.20 GeV/c
 obtained due to specific cuts.}
\label{fig:Fig5}
\end{figure}

  But the diagrams shown in Fig.3 are insufficient to describe totally the characteristics of the 
reaction $np \rightarrow np \pi^+ \pi^-$. It is necessary to take into account the diagrams of 
the following type:
\begin{figure}[h]
\includegraphics[width=0.75\textwidth]{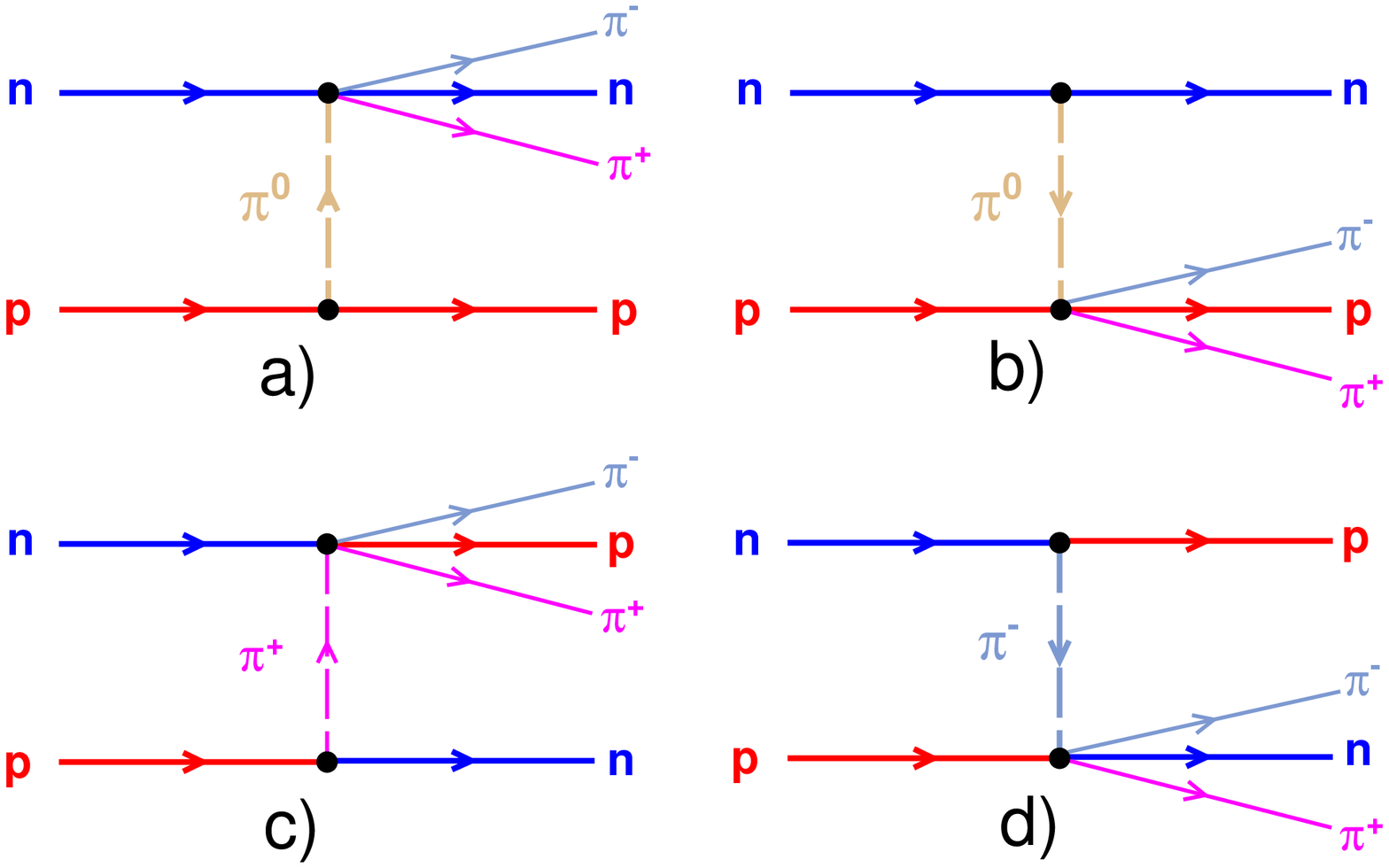}
\caption{OPER diagrams 1$\times$3 for the reaction $np \rightarrow np \pi^+ \pi^-$}
\label{fig:Fig6}
\end{figure}

 the matrix element  for which is written in the following form:
 $$M_3 = G \bar u(q_N)\gamma_5 u(Q_N)·F_1·T_{\pi N \rightarrow \pi \pi N}/(t - m_{\pi}^2),$$
  where $T_{\pi N \rightarrow \pi \pi N}$  - off mass shell amplitudes of inelastic 
  $\pi N \rightarrow \pi \pi N$ - scattering
  that are known much worse than elastic $T_{\pi N \rightarrow \pi N}$ amplitudes.
  Therefore it is necessary to do a parametrization of the inelastic 
  $\pi N \rightarrow \pi \pi N$-scattering (see Appendix).
\begin{figure}[h]
\includegraphics[width=0.85\textwidth]{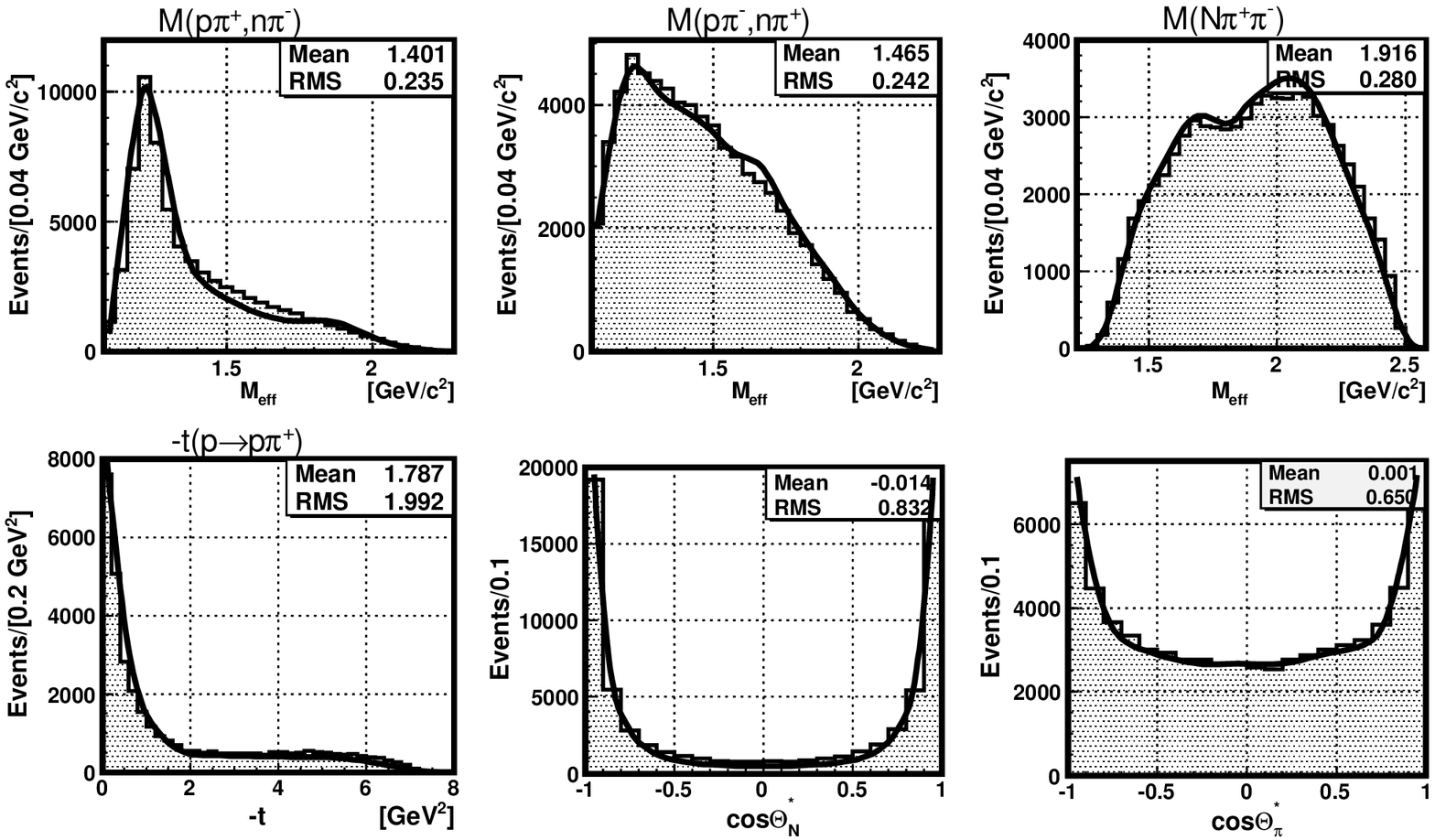}
\caption{Distributions for the reaction $np \rightarrow np \pi^+ \pi^-$ at $P_0$=5.20 GeV/c}
\label{fig:Fig7}
\end{figure}

  It permits to get a good description of the experimental characteristics of the reaction 
 $np \rightarrow np \pi^+ \pi^-$ at $P_0$=5.20 GeV/c (Fig.7) taking into account
OPER~diagrams shown in Fig.3 and Fig.6 :

\section{The reaction $\bar pp \rightarrow \bar pp \pi^+ \pi^-$ at $P_0$ = 7.23 GeV/c}

  Using OPER~model we try to describe the experimental distributions from the reaction
$\bar pp \rightarrow \bar pp \pi^+ \pi^-$ at $P_0$ = 7.23 GeV/c~\cite{pbarp} 

 It is observed a good agreement between experimental data and theory in Fig.8.                                                              

\begin{figure}[h]
\includegraphics[width=0.9\textwidth]{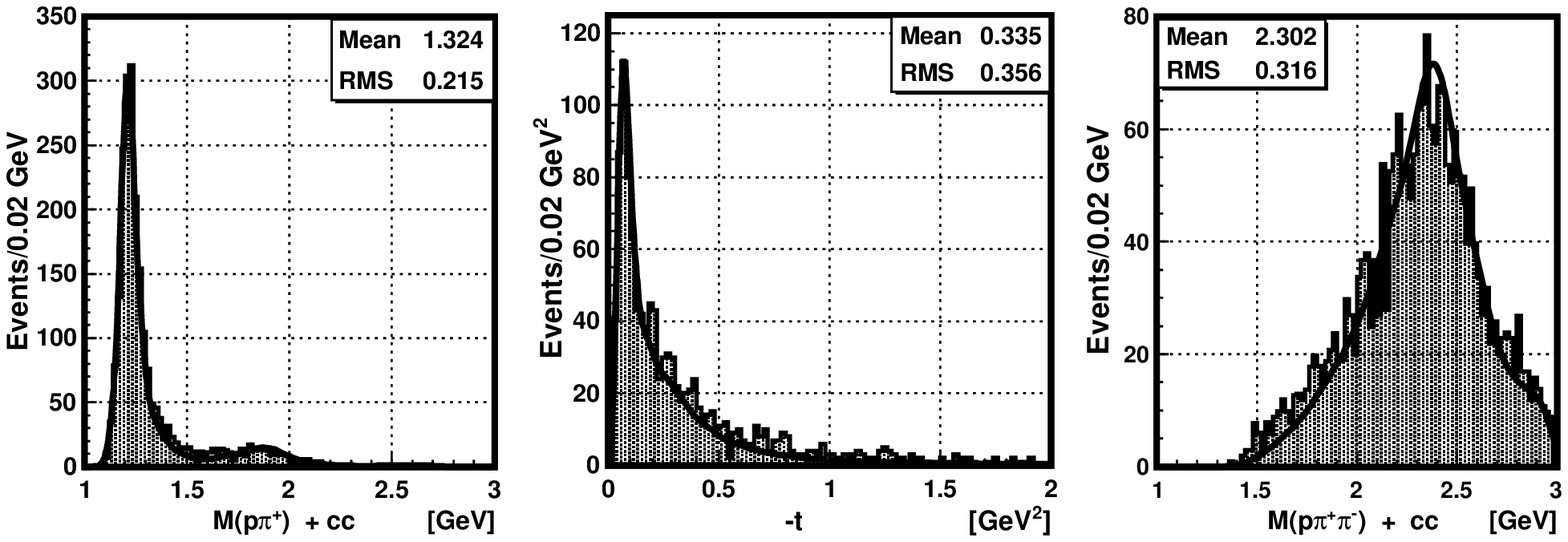}
\caption{Distributions for the reaction $\bar pp \rightarrow \bar pp \pi^+ \pi^-$ 
 at $P_0$ = 7.23 GeV/c}
\label{fig:Fig8}
\end{figure}
\section{The reaction $np \rightarrow np \pi^+ \pi^-$ at $P_0$ $<$ 3 GeV/c}
  The study of effective mass spectra of $np$ - combinations at $P_0$=1.73 and 2.23 GeV/c 
(Fig.9) shows the clear peack close the threshold ($M_{np}= m_n+m_p$) that can not be described
within the framework of OPER-model using the diagrams from Fig.3 and Fig.6.
\begin{figure}[h]
\includegraphics[width=0.9\textwidth]{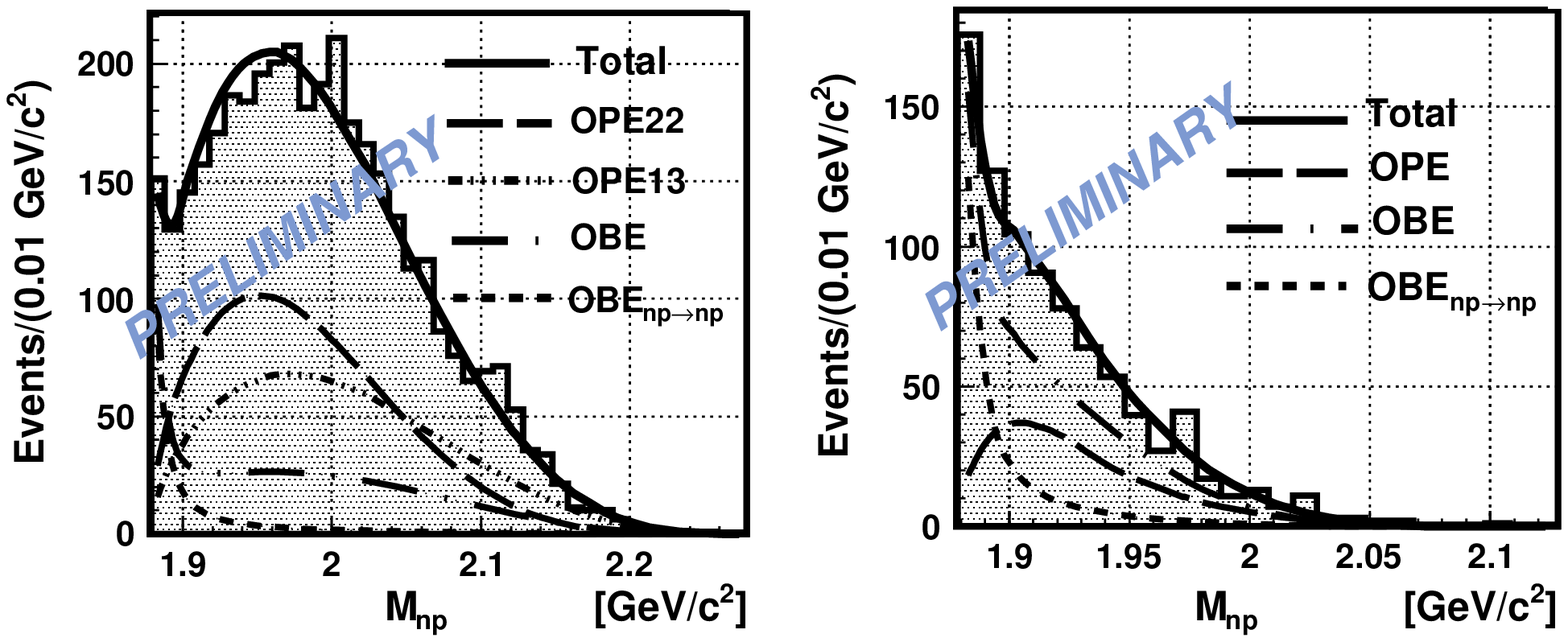}
\caption{The distributions of $M_{np}$ for treaction $np \rightarrow np \pi^+ \pi^-$ 
at $P_0$ = 2.23 GeV/c (left) and 1.73 GeV/c (right).}
\label{fig:Fig9}
\end{figure}

  The model of Regge poles with baryon exchange and nonlinear trajectories, suggested in~\cite{OBE}
was used to describe these features.
The following diagrams of one baryon exchange (OBE) were taken into account  within the
framework of this model:
\begin{figure}[h]
\includegraphics[width=0.85\textwidth]{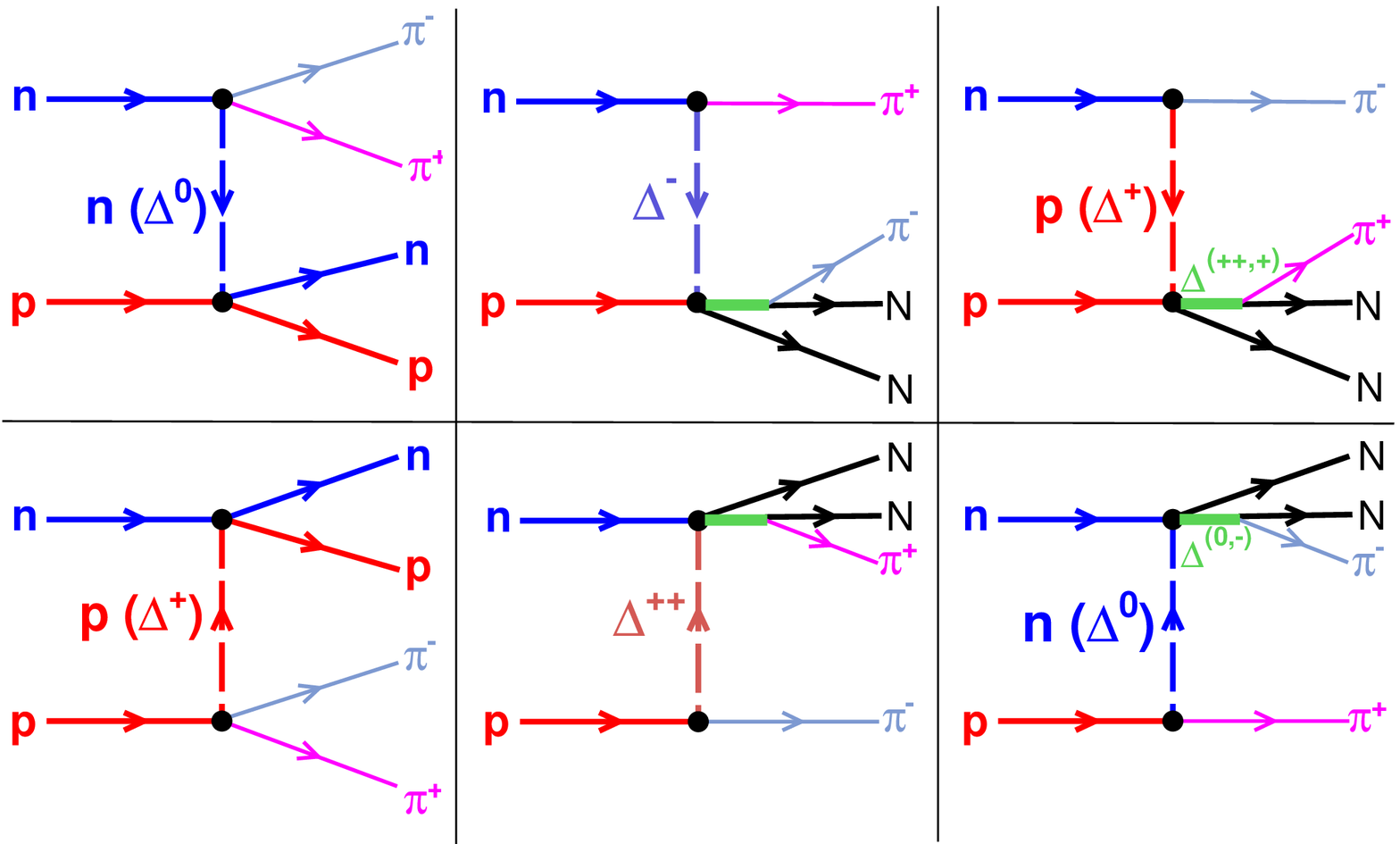}
\caption{OBE diagrams for the reaction $np \rightarrow np \pi^+ \pi^-$}
\label{fig:Fig10}
\end{figure}

   The vertex function of elastic $np \rightarrow np$ scattering was calculated using the data 
from~\cite{NNel}. 

   The vertex functions of $\Delta N \rightarrow n p$, $N N \rightarrow \Delta N$ and 
$\Delta N \rightarrow \Delta N$ scattering were calculated corresponding to~\cite{NNND}.
In result  one can get the good description of the experimental distribution from the
reaction $np \rightarrow np \pi^+ \pi^-$ at $P_0$ = 1.73 and 2.23  GeV/c (Fig.9 and Fig.11).
\vspace{-0.1cm}
\begin{figure}[h]
\includegraphics[width=0.9\textwidth]{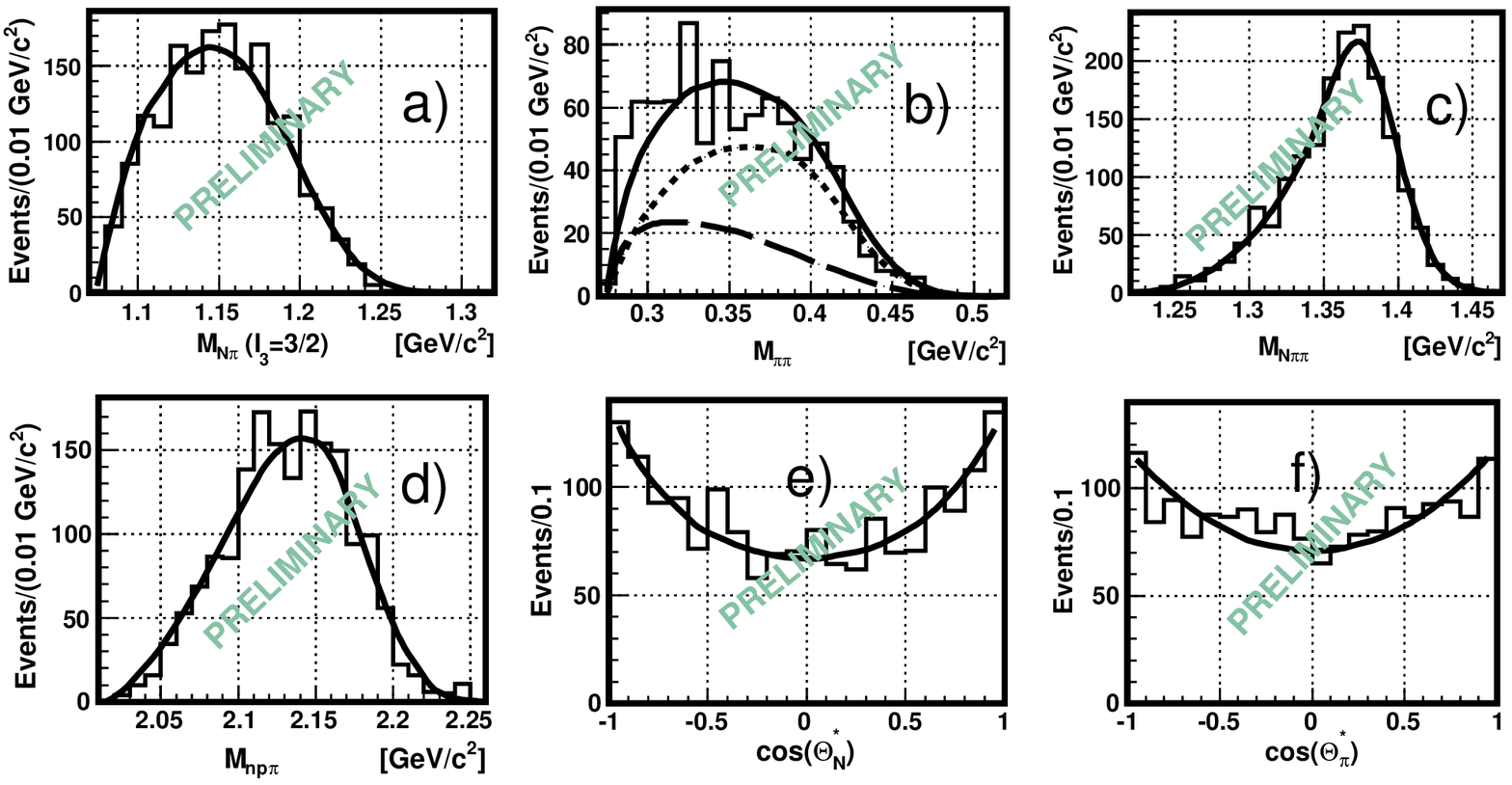}
\caption{Distributions for the reaction $np \rightarrow np \pi^+ \pi^-$ at $P_0$=1.73 GeV/c}
\label{fig:Fig11}
\end{figure}

\section{OPER model and other reactions}

 The other reactions of np interactions are scheduled to study by means of OPER~model:\\
\hspace*{1.5cm} $n p \rightarrow p p \pi^-$ 
\hspace*{1.55cm} vertex functions 1$\times$2\\
\hspace*{1.5cm} $n p \rightarrow p p \pi^- \pi^0$ 
\hspace*{1.15cm} vertex functions 2$\times$2 and 1$\times$3\\
\hspace*{1.5cm} $n p \rightarrow p p \pi^+ \pi^- \pi^-$ 
\hspace*{0.55cm} vertex functions 2$\times$3\\ 
\hspace*{1.5cm} $n p \rightarrow p p \pi^+ \pi^- \pi^- \pi^0$ 
\hspace*{0.15cm} vertex functions 3$\times$3\\  
\hspace*{1.5cm} $n p \rightarrow n p \pi^+ \pi^- \pi^+ \pi^-$ 
\hspace*{0.05cm} vertex functions 3$\times$3\\

   Similar reactions of $pp$, $\bar p p$ and $\pi N$ interactions also can be described 
by OPER~model.
The following reactions were simulated for HADES experiment:\\
\hspace*{1.5cm}  $pp \rightarrow pp \pi^+ \pi^-$ at $T_{kin}$=3.5 GeV\\
\hspace*{1.5cm}  $np \rightarrow np \pi^+ \pi^-$ at $T_{kin}$=1.25 GeV\\
\hspace*{1.5cm}  $np \rightarrow np e^+ e^-$     at $T_{kin}$=1.25 GeV with vertex function
   of $\gamma N \rightarrow N e+ e^-$.\\

   Since the $\pi N \rightarrow \pi N$  and $\pi N \rightarrow \pi \pi N$ vertex functions 
 are taken in helicity representation it seems to be perspective to use OPER~model for description
 of the reaction with polarized particles.

\section{Conclusion}

  Reaction $np \rightarrow np \pi^+ \pi^-$ is characterized by the plentiful production of 
the $\Delta$ resonance and the large peripherality of the secondary particles. The experimental 
data are successfully described by the further development of OPER~model.\\

 However  at  $P_0$ $<$ 3 GeV/c it is necessary to take into account another mechanism of the
reaction (such as OBE).\\

  OPER~model permits to describe another $N(\bar N)-N$ reactions with the production of
some $\pi$-mesons. The further development of OPER-model  can be very promising to describe the 
production of $e^+ e^-$-pairs in hadronic interactions.\\

  OPER~model can be used as an effective tool to simulate various reactions of hadronic
interactions.

\vspace{1.5cm}
\section{Appendix: Parametrization of $\pi N \rightarrow \pi \pi N$ reactions}

  Within the framework of Generalized Isobar Model (GIM)~\cite{GIM} $\pi N \rightarrow \pi \pi N$ 
reactions are described as quasi-two body ones ($a + b \rightarrow c + d$):\\ 
\hspace*{1.5cm} $\pi N \rightarrow N^*(\Delta^*) \rightarrow \Delta \pi$,\\
\hspace*{1.5cm} $\pi N \rightarrow N^*(\Delta^*) \rightarrow N \rho$\\
\hspace*{1.5cm} $\pi N \rightarrow N^*(\Delta^*) \rightarrow N \epsilon$\\ 
\hspace*{1.5cm} $\pi N \rightarrow N^*(\Delta^*) \rightarrow N^*_{1440} \pi$

with the consequent decays:\\
\hspace*{2.5cm} $\Delta \rightarrow N \pi$,\\
\hspace*{2.5cm} $\rho \rightarrow \pi \pi$,\\
\hspace*{2.5cm} $\rho \rightarrow \pi \pi$,\\
\hspace*{2.5cm} $N^*_{1440} \rightarrow N \pi$\\

 The parameters of the following resonances (****  and ***) were taken from 
 Review~of~Particle~Properties:
$$
 \begin{array}{cc}
 N^*(1440) P11 & D^*(1600) P33\\
 N^*(1520) D13 & D^*(1620) S31\\
 N^*(1675) D15 & D^*(1700) D33\\
 N^*(1680) F15 & D^*(1900) S31\\
 N^*(1720) P13 & D^*(1905) F35\\
 N^*(2000) F15 & D^*(1910) P31\\
 N^*(2080) D13 & D^*(1920) P33\\
 N^*(2190) G17 & D^*(1940) D33\\
               & D^*(1950) F37
 \end{array}
$$                                         
     The spin and isospin relations were taken account.\\

  For quasi two-body reactions like $a + b \rightarrow c + d$ one can write
 $$d\sigma = \frac{1}{(2S_a+1)(2S_b+1)} \left(\frac{2\pi}{p}\right)^2 
 \sum_{\lambda_i}|<\lambda_d\lambda_c |T|\lambda_b\lambda_a>|^2 \times dPS\, ,$$
 $$<\lambda_d \lambda_c |T| \lambda_b \lambda_a> = \frac{1}{4\pi} \sum_j (2j+1)
   <\lambda_d \lambda_c |T_j| \lambda_b \lambda_a>  e^{i(\lambda - \mu) \varphi}
   d^j_{\lambda \mu}(\theta) \, .$$
where
 $\lambda=\lambda_a- \lambda_b$, $\mu=\lambda_c - \lambda_d$ - helicity 
 variables,\\
\hspace*{1.0cm} $d^j_{\lambda \mu}(\theta)$ - rotation matrixe,\\
\hspace*{1.0cm} $dPS$ - phase space element.\\ 

 The polarization components of the particles c and d from the reaction 
 $a + b \rightarrow c + d$ 
is suiteable to express through the elements of the spin density matrix 
(for example, for particle~d):
 $$\rho^d_{mm\prime} = \frac{1}{N} \sum_{\lambda_c\lambda_b\lambda_a}
   <m\prime\lambda_c|T|\lambda_b\lambda_a>^*<m\lambda_c|T|\lambda_b\lambda_a>$$
where  normalization factor for Sp$\rho$=1:
$$N = \sum_{m\lambda_c\lambda_b\lambda_a} <m\lambda_c|T|\lambda_b\lambda_a>^2.$$

\newpage
\underline{\bf Example}:\\ 
 $$\pi + N \rightarrow N^*_{1680} \rightarrow \Delta + \pi \rightarrow 
   (N + \pi) + \pi$$ 
 $$<\lambda_{\Delta}|T|\lambda_N> = 
   C^{\frac{5}{2},-\lambda_N}_{3,0;\frac{1}{2},-\lambda_{\Delta}}
   C^{\frac{5}{2},-\lambda_{\Delta}}_{1,0;\frac{3}{2},-\lambda_{\Delta}}
   d^{\frac{5}{2}}_{-\lambda_N,-\lambda_{\Delta}}(\theta) \times R_J,$$
 where $R_J$ is taken in Breight-Wigner form.\\

    Then it is easy to get the angular distribution of $\Delta$ (in CMS): 
 $$\frac{d\sigma(s,t)}{d\Omega} \sim (1+2\cos^2\theta_{\Delta})\mid R_J\mid^2=
   (1+2\cos^2\theta_{\Delta}) \times BW(\sqrt{s},M_R,\Gamma_R)$$
\\
If particle d is unstable: $d \rightarrow \alpha + \beta$ 
$(d \rightarrow \Delta + \pi)$ then in the rest system of the particle d:
 $$W_{\Delta} = \frac{3}{4\pi} \Bigl \{ \rho_{33}\sin^2\theta+
   \frac{1}{3}\rho_{11}(1+3\cos^2\theta)-
   \frac{2}{\sqrt{3}}Re\rho_{3-1}\sin^2\theta\cos2\varphi-
   \frac{2}{\sqrt{3}}Re\rho_{31}\sin2\theta\cos\varphi \Bigr \}$$
  - is the normalized angular distribution of the decay products.\\

 To compare with experimental data the following cross-sections were calculated 
using  GIM (Fig.12):
\begin{figure}[h]
\includegraphics[width=1.0\textwidth]{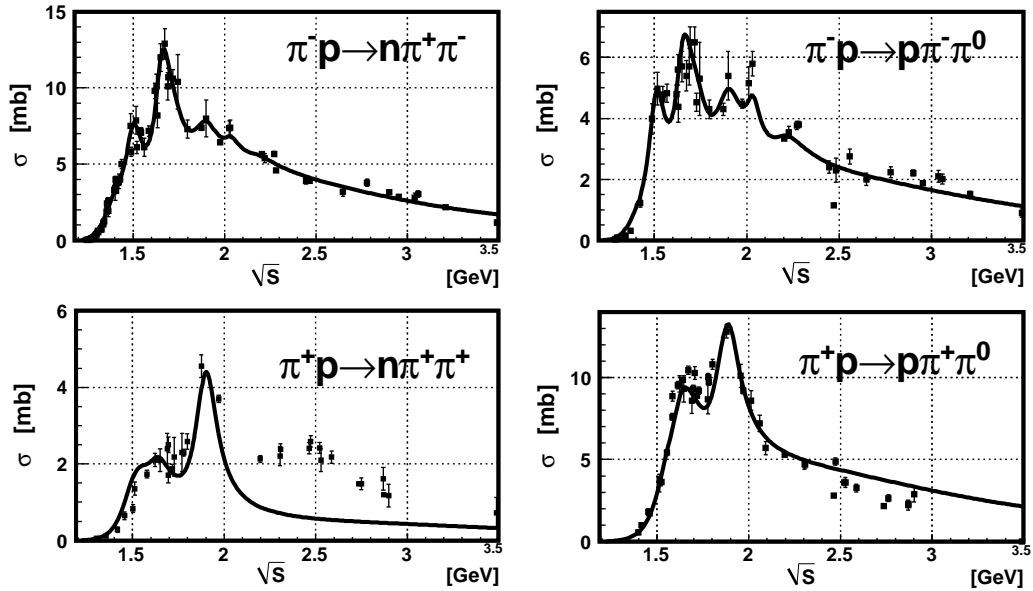}
\caption{Cross-sections of the $\pi n \rightarrow \pi \pi N$ reactions.}
\label{fig:Fig12}
\end{figure}

\vspace{0.5cm}
 One can see a satisfactory description of cross-sections, except 
$\pi^+ p \rightarrow n \pi^+ \pi^+$.
May be it is necessary to take into account S-wave of $\pi^+ \pi^+$ scattering 
with I=2 in GIM.\\ 

Some distributions of the reaction $\pi^- p \rightarrow  n \pi^+ \pi^-$ were 
calculated at various energies to study a quality of the application of GIM 
(Fig.13):\\ 
\\
\begin{figure}[h]
\includegraphics[width=0.9\textwidth]{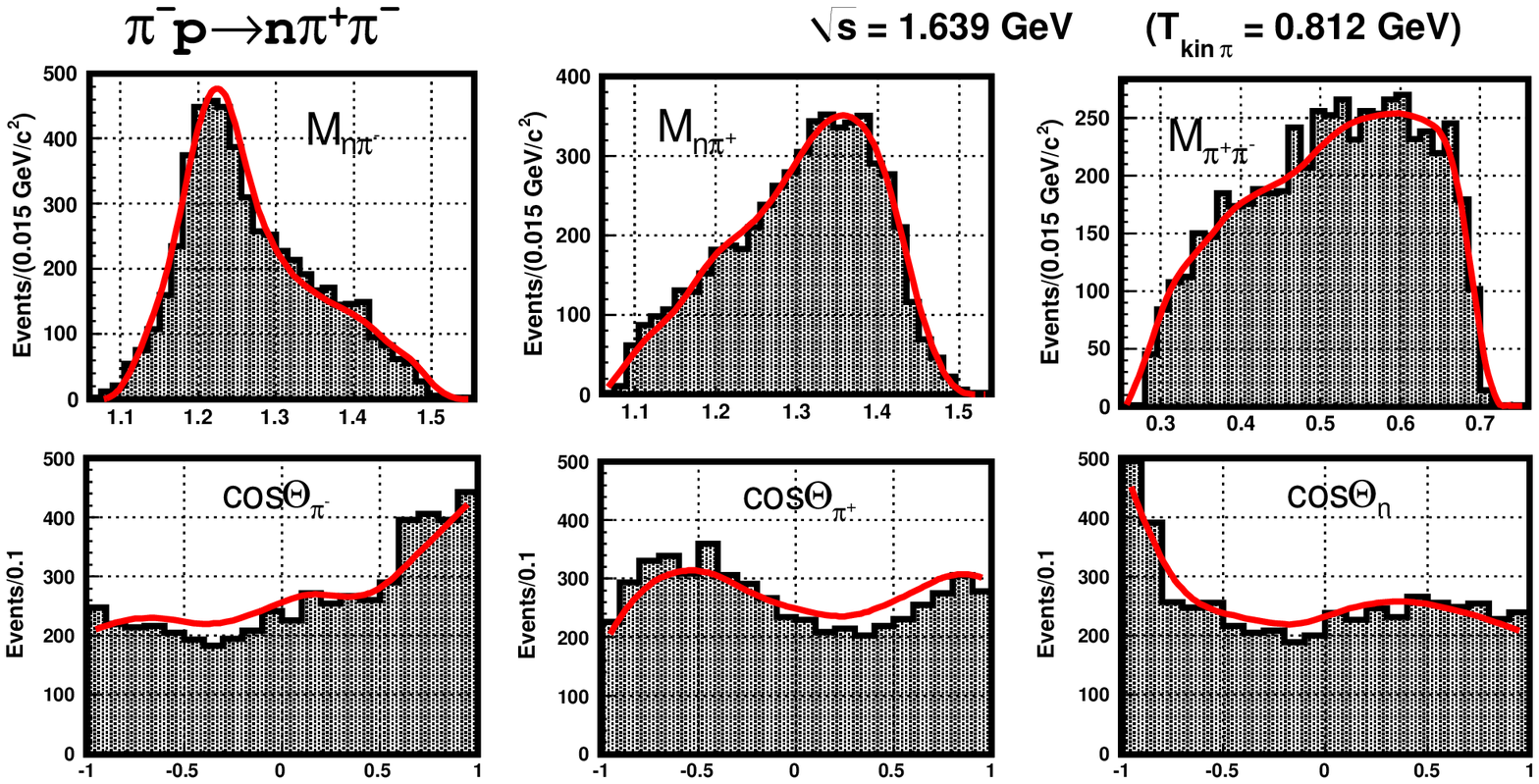}
\caption{Some distributions from the reaction $\pi^- p \rightarrow \pi^+ \pi^- n$
 at $T_{kin}$=1.0 GeV~\cite{Dolbeau}}
\label{fig:Fig13}
\end{figure}

It is observed a good agreement between experimental data and theory.




\end{document}